\documentclass{article}
\usepackage{graphics}

\begin{document}

\title{An Algorithm for Clock Synchronization with the Gradient Property in Sensor Networks}

\author{Rodolfo~M.~Pussente\\
Valmir~C.~Barbosa\thanks{Corresponding author (valmir@cos.ufrj.br).}\\
\\
Universidade Federal do Rio de Janeiro\\
Programa de Engenharia de Sistemas e Computa\c c\~ao, COPPE\\
Caixa Postal 68511\\
21941-972 Rio de Janeiro - RJ, Brazil}

\date{}

\maketitle

\begin{abstract}
We introduce a distributed algorithm for clock synchronization in sensor
networks. Our algorithm assumes that nodes in the network only know their
immediate neighborhoods and an upper bound on the network's diameter.
Clock-synchronization messages are only sent as part of the communication,
assumed reasonably frequent, that already takes place among nodes. The algorithm
has the gradient property of \cite{fl06}, achieving an $O(1)$ worst-case skew
between the logical clocks of neighbors. As in the case of \cite{lms85,wl88},
the algorithm's actions are such that no constant lower bound exists on the rate
at which logical clocks progress in time, and for this reason the lower bound of
\cite{fl06,mt05} that forbids constant skew between neighbors does not apply.

\end{abstract}

\section{Introduction}

We consider a network of sensors and assume it may be represented by a connected
undirected graph $G=(N,E)$ whose nodes stand for sensors and undirected edges
for bidirectional communication channels. We also assume that channels are fully
reliable and deliver messages with delays bounded by a constant. We let
$n=\vert N\vert$, use $N_i\subset N$ to denote the set of node $i$'s neighbors,
and $t\ge 0$ to denote real time.

No node has access to the value of $t$ but rather relies on a hardware clock to
estimate it. For node $i$, the hardware clock at time $t$ is denoted by
$H_i(t)\ge 0$. Ideally, $H_i(t)$ should evolve in ``lockstep'' with $t$, but we
assume instead that its progress occurs at a positive rate that may drift as $t$
elapses. We assume an additive drift, which at time $t$ is denoted by
$\rho_i(t)\in[-\hat\rho,\hat\rho]$ for some constant $\hat\rho\in[0,1)$. The
rate at which $H_i(t)$ progresses is then $1+\rho(t)$ at time $t$, and it
follows that
\begin{equation}
H_i(t)=\int_{r=0}^t[1+\rho(r)]\,dr.
\end{equation}

Because the instantaneous drifts may be different throughout $G$ for any given
$t$, nodes may only acquire a common estimate of real time by resorting to clock
synchronization. At node $i$, this amounts to maintaining a logical clock
$L_i(t)\ge 0$ that normally progresses at a rate proportional to that of the
node's hardware clock but can be updated as $i$ learns about the logical clocks
of other nodes in $G$.

We assume that $L_i(t)$ is never allowed to run backwards (i.e.,
$L_i(t')\ge L_i(t)$ for all $t'>t$), and note that this is sometimes made more
stringent by requiring a constant lower bound $b\in(0,1]$ on the rate of
progress of every node's logical clock.\footnote{As in \cite{st87,ege02}, but
not in \cite{lms85,wl88}, for example.} When the latter is the case, enforcing
the requirement is easy if $\hat\rho$ is known to the nodes: it suffices to set
$dL_i(t)/dH_i(t)\ge b/(1-\hat\rho)$, since
\begin{equation}
\frac{dL_i(t)}{dt}=\frac{dL_i(t)}{dH_i(t)}\frac{dH_i(t)}{dt}\ge
\frac{b}{1-\hat\rho}(1-\hat\rho)=b.
\end{equation}

The goal of a distributed algorithm for clock synchronization is to minimize the
skew $\vert L_i(t)-L_j(t)\vert$ for all pairs $i,j$ of distinct nodes and all
$t$. While significant progress was achieved in the past (cf., e.g.,
\cite{lms85,st87,wl88,ege02}), with a single exception to be discussed shortly
it seems that all algorithms to date admit a worst-case skew of $O(D)$, where
$D$ is the diameter of $G$, even between neighbors in the graph. The problem
with this in the context of sensor networks is that, for tasks as fundamental as
that of data fusion \cite{qwic01}, for example, nearby nodes must synchronize
their clocks much more strictly than this, while for distant nodes the larger
skew is not a problem.

This observation has motivated the introduction in \cite{fl06} of a new
property of clock skews, the so-called gradient property. For $f$ a positive,
nondecreasing real function of distances in $G$, and $d_{ij}$ the distance
between nodes $i$ and $j$, the gradient property requires
\begin{equation}
\vert L_i(t)-L_j(t)\vert\le f(d_{ij})
\end{equation}
for all pairs $i,j$ of distinct nodes and all $t$. To our knowledge, the only
algorithm to date that guarantees clock skews for which the gradient property
holds is the one of \cite{lw06}. In this algorithm, we have that
\begin{equation}
f(d_{ij})\textrm{ is }\cases{
O(d_{ij}\sqrt{D}),&if $d_{ij}\le\sqrt{D+1}$;\cr
O(D),&otherwise,\cr
}
\end{equation}
so in the worst case the clock skew between neighbors in $G$ is $O(\sqrt{D})$.

Achieving this, however, requires a relatively strong assumption on what is
known to the nodes and also that nodes communicate frequently with their
neighbors. The assumption is that both $D$, the graph's diameter, and
$\hat\rho$, the maximum drift of hardware-clock rates, are known to all nodes.
As for communicating with neighbors, a node is required to do so whenever its
logical clock reaches a new integer value or is updated in the wake of the
reception of a message.

While for some sensor networks the assumption may be regarded as reasonable,
since it may be possible to bound both $D$ and $\hat\rho$ from above in the
environment in question, we find the need for frequent communication with
neighbors to be generally incompatible with the power-consumption constraints
normally associated with sensor networks. So we maintain the assumption, in
part, but strive to reduce communication requirements as much as possible.

\section{A new algorithm}

Unlike the algorithm of \cite{lw06}, the algorithm we introduce in this paper
targets sensor networks directly. For this reason, we adopt the same two
assumptions as \cite{mt05} regarding the communication among sensors:
\begin{enumerate}
\item[(i)] Messages sent between neighbors in $G$ are delivered instantaneously;
\item[(ii)] If $t$ and $t'$ are instants at which two neighbors communicate in
one of the two directions without any intervening communication in the same
direction between them in the meantime, then $\vert t-t'\vert\le d$ for some
$d>0$.
\end{enumerate}
We aim at synchronizing clocks without any messages sent exclusively for this
purpose, that is, by attaching clock-synchronization messages to whatever
communication is already guaranteed to take place by assumption (ii).

We assume that nodes know their local neighborhoods (i.e., the neighbor set
$N_i$ for node $i$) and, like \cite{lw06}, that the diameter $D$ (or an upper
bound on it) is also known to them. We assume further that no node has access to
the value of $\hat\rho$ or $d$, and that clock synchronization is started
concurrently at any number of nodes, from which it propagates. If $i$ is one of
these nodes, then we assume $L_i(t)=0$ for $t$ the time at which clock
synchronization is started at node $i$; if not, then we assume $L_i(t)=0$ for
$t$ the time at which $i$ is first reached by a clock-synchronization message.

For $j\in N_i$, node $i$ maintains a variable $L_i^j$ to store its current view
of the logical clock of $j$. If $t$ is the instant at which $L_j(t)$ is
communicated by $j$ to $i$, and if $L_i^j$ results from this communication, then
assumption (i) implies that $L_i^j=L_j(t)$. For all $i\in N$ and all $j\in N_i$,
we assume $L_i^j=0$ before the reception at $i$ of the first
clock-synchronization message from $j$.

Now let $\alpha_i$ be the number by which the current rate of progress of
$L_i(t)$ is proportional to that of $H_i(t)$; that is, let
$\alpha_i=dL_i(t)/dH_i(t)$. Our algorithm uses $\alpha_i$ as the minimum of
multiple $\alpha_i^j$'s, one for each of node $i$'s neighbors, that is,
\begin{equation}
\alpha_i=\min_{j\in N_i}\alpha_i^j.
\end{equation}
We assume that, initially, $\alpha_i^j=1$ for all $i\in N$ and all $j\in N_i$.
Our algorithm is based on lowering $\alpha_i$ as needed whenever node $i$
detects, upon receiving a clock-synchronization message, that its logical clock
is ahead of that of the message's sender by a certain amount $c$ or more.

Other than this manipulation of $\alpha_i$, our algorithm strives at node $i$ to
advance $L_i(t)$, if appropriate, toward the greatest $L_i^j$, so long as this
does not leave the least $L_i^j$ behind by the same $c$ as above or more. We
now describe our algorithm in terms of how node $i$ responds to the reception of
$\langle L\rangle$ in a clock-synchronization message from node $j\in N_i$ at
time $t$. Notice that, by assumption (i), $L=L_j(t)$. Node $i$'s response to the
message from $j$ comprises the following two steps, whose processing is also
assumed instantaneous.

\begin{description}
\item[Step 1.] $L_i^j:=L$.
\item[Step 2.] With $L^-=\min_{j\in N_i}L_i^j$ and $L^+=\max_{j\in N_i}L_i^j$:
\begin{enumerate}
\item[(a)] If $L_i(t)\ge L_i^j+c$, then $\alpha_i^j:=1/D$, otherwise
$\alpha_i^j:=1$.
\item[(b)] $L_i(t):=\max\{L_i(t),\min\{L^-+c,L^+\}\}$.
\end{enumerate}
\end{description}

Step~1 is devoted simply to updating node $i$'s view of node $j$'s logical
clock. Step~2 attempts to reduce $\alpha_i^j$ to $1/D$, in case
$L_i(t)\ge L_i^j+c$ (and thus $L_i(t)\ge L^-+c$); or to restore $\alpha_i^j$ to
$1$, in case $L_i(t)<L_i^j+c$; or yet to advance $L_i(t)$, in case both
$L_i(t)<L^-+c$ and $L_i(t)<L^+$.

The value of $L_i(t)$ that results from Steps~1 and~2 continues to evolve before
it gets sent in a clock-synchronization message to some of $i$'s neighbors. If
such a message is sent at some time $t'$ before $i$ receives the next
clock-synchronization message, then $\langle L_i(t')\rangle$ gets sent along
with it such that $L_i(t')\le L_i(t)+\alpha_i(1+\hat\rho)d$.

\section{Worst-case clock skews}

Step~2(b), with $c=(1+\hat\rho)\sqrt{D+1}$, is the essence of the algorithm in
\cite{lw06}. The reason why that algorithm guarantees a maximum skew of
$O(\sqrt{D})$ between the logical clocks of neighbors in $G$ is intimately
related to this particular choice for $c$ and to how this choice relates to the
worst-case skew between any two nodes, which is always less than
$(1+\hat\rho)D+1$. There are other factors involved, but this one is crucial and
a closer examination of \cite{lw06} reveals that choosing $c$ to be $O(1)$, for
example, disrupts the clocks' gradient property.

Considered within the assumptions of our model, the problem with letting $c$ be
$O(1)$ in Step~2(b) is that a length-$O(D)$ wait chain may occur in $G$ in
which each node finds out that its logical clock is ahead of the next node's by
at least $c$. In this chain, the node whose logical clock is ahead of all
others' may dart still farther ahead unchecked for an $O(D)$ amount of time,
which will then be the worst-case skew between neighbors. So, in order to
accommodate the possibility of a constant value for $c$ along with the gradient
property for some $f$, a mechanism is needed to slow down the progress of
logical clocks that are ahead of others by $c$ or more. This is what Step~2(a)
does, provided $c\le(1+\hat\rho)d$, as we assume henceforth. As we demonstrate
shortly, an $f$ is achieved that implies constant skew between the logical
clocks of neighbors.

Let us now examine the worst-case skews that logical clocks may have under
Steps~1 and~2. We start with the skew between any two nodes, in which case it
suffices that we consider a chain of $D+1$ nodes and the algorithm's start-up
process. After the algorithm is initiated by one of the nodes (this gives us the
worst case as far as the number of initiators is concerned), it may take as long
as $Dd$ time units for all others to have started their logical clocks, during
which time the initiator may advance its logical clock from $0$ to at most
$(1+\hat\rho)Dd$. This is then the largest skew between any two logical clocks.

We now turn to the worst-case skew between the logical clocks of neighbors in
$G$. As we indicated above, Step~2(a) has a crucial role to play in ensuring
that this skew remains bounded within the desired limits of $O(1)$. In order to
see that this is really the case, first recall that, in the absence of
Step~2(a), Step~2 would be ineffectual at time $t$ if we had
$L_i(t)\ge L_j(t)+c$. The following, then, is fundamentally dependent on
Step~2(a).

Let us consider the same $(D+1)$-node chain as above and look at the situation
in which $L_i(t)=L_j(t)+c$, $L_j(t)=L_k(t)+c$ for some $k\neq i$, and so on
through the chain. Clearly, this scenario can only involve so many edges of the
chain. If we let $\ell$ be this number of edges, then our previous result on the
maximum skew between any two nodes, together with the fact that
$c\le(1+\hat\rho)d$, implies that
\begin{equation}
\ell=\min\left\{D,\frac{(1+\hat\rho)Dd}{c}\right\}=D.
\end{equation}
By time $t+d$, each of the first $\ell$ nodes in the chain ($i$, $j$, and so on)
has found out that it is waiting for its neighbor down the chain to catch up
with it, and consequently has reduced its rate to $1/D$. The $\ell+1$st node
has caught up with its predecessor, but $j$ will not be able to catch up with
$i$ for another $(\ell-1)d$ time units, every $d$ of which sees a new node
ready to raise its rate back to $1$ and catch up with its own predecessor.

During the first $d$ time units past time $t$, node $i$'s logical clock may
increase by as much as $\alpha_i(1+\hat\rho)d\le(1+\hat\rho)d$, node $j$'s by
as little as $\alpha_j(1-\hat\rho)d\ge(1-\hat\rho)d/D$, thus causing the logical
clocks of $i$ and $j$ to undergo a further separation of at most
\begin{equation}
(1+\hat\rho)d-\frac{(1-\hat\rho)d}{D}\le(1+\hat\rho)d.
\end{equation}
During the remaining $(\ell-1)d$ time units, the logical clock of node $i$ may
increase by as much as
\begin{equation}
\alpha_i(1+\hat\rho)(\ell-1)d=\frac{(1+\hat\rho)(\ell-1)d}{D}.
\end{equation}
The logical clock of node $j$, in turn, may during this time increase by as
little as
\begin{equation}
\alpha_j(1-\hat\rho)(\ell-1)d=\frac{(1-\hat\rho)(\ell-1)d}{D}.
\end{equation}
At time $t+\ell d$, then, the greatest possible skew between the logical clocks
of the neighboring nodes $i$ and $j$ is
\begin{equation}
c+(1+\hat\rho)d+\frac{2\hat\rho(\ell-1)d}{D}\le c+(1+3\hat\rho)d,
\end{equation}
since $\ell=D$. Our algorithm is then seen to achieve the gradient property in
such a way that
\begin{equation}
f(d_{ij})\textrm{ is }O(d_{ij})
\end{equation}
for all $d_{ij}\in[1,D]$, so the worst-case clock skew between neighbors is
$O(1)$.

\section{Discussion}

In \cite{fl06}, and also in \cite{mt05} for the specific case of assumptions (i)
and (ii), it is proven that $f(d_{ij})$ is $\Omega(d_{ij}+\log D/\log\log D)$.
This is proven for all clock-synchronization algorithms that have the gradient
property and for which the constant lower bound $b$ mentioned earlier on the
rate of progress of all logical clocks exists. Such a property would be
seriously at odds with our claim of an $O(1)$ worst-case skew between neighbors,
so in this section we discuss its relation to our algorithm. Specifically, we
demonstrate that our approach admits no constant lower bound on $dL_i(t)/dt$
that holds for all $i$ and all $t$, so the lower bound on $f(d_{ij})$ does not
hold.

We first discuss the definability of $dL_i(t)/dt$. For fixed $t$, let $t_1<t$
and $t_2>t$ be such that the value of $\alpha_i$ does not change in the time
interval $[t_1,t)$ or in the interval $(t,t_2]$. Then $dL_i(t)/dt$ is in
principle definable indistinctly as
\begin{equation}
\lim_{t_1\to t}\frac{L_i(t)-L_i(t_1)}{t-t_1}
=\alpha_i\lim_{t_1\to t}\frac{H_i(t)-H_i(t_1)}{t-t_1}
=\alpha_i\frac{dH_i(t)}{dt}
\end{equation}
or
\begin{equation}
\lim_{t_2\to t}\frac{L_i(t_2)-L_i(t)}{t_2-t}
=\alpha_i\lim_{t_2\to t}\frac{H_i(t_2)-H_i(t)}{t_2-t}
=\alpha_i\frac{dH_i(t)}{dt}.
\end{equation}
However, if $t$ is precisely the time at which node $i$ changes the value of
$\alpha_i$ through Step~2(a) (by reducing some $\alpha_i^j$ from $1$ to $1/D$
while all others remain equal to $1$, or by raising the single $\alpha_i^j$
whose value is $1/D$ back to $1$), then the two limits above are inconsistent
with each other and $dL_i(t)/dt$ remains undefined.

But the values of $t$ for which $dL_i(t)/dt$ is undefined are only finitely
many, so one naturally wonders about the other, infinitely many instants at
which the derivatives are defined. For these other instants, notice that
Step~2(a) never causes $\alpha_i$ to be reduced below $1/D$, so one might still
consider, for all $i\in N$, the existence of the lower bound $b$ on
$dL_i(t)/dt$, provided
\begin{equation}
b\le\frac{(1-\hat\rho)}{D}.
\end{equation}
Such a bound, however, would not be a constant, as it would depend on $G$.

\section{Concluding remarks}

Our algorithm's Step~2 embodies two competing trends in its two parts (a) and
(b). The aim of part (a) is to slow down nodes whose logical clocks are ahead of
any of their neighbors' by $c$ or more. Part (b), on the other hand, forces a
node's logical clock to move ahead toward its neighbors' whenever possible. Both
trends are fundamental to the algorithm's proper operation. Without Step~2(a),
the $O(1)$ worst-case skew between neighbors would be unachievable; without
Step~2(b), the presence of a single slow-moving hardware clock would slow down
all nodes' logical clocks, turning them into poor approximations of real time.

One relevant open question at this point is how the two trends balance each
other, both in theory and in practice. Our algorithm relies strongly on the
possibility of altering, in Step~2(a), the rates at which nodes' logical clocks
follow their hardware clocks. Even though there is a clear provision for such
rates to return to their original value of $1$ whenever safe, further
investigation is needed to clarify their most important properties. One of these
concerns the duration of the periods during which the rates get reduced. Another
is related to how rate reduction affects the logical clocks' main purpose, which
is to track the progress of real time in as synchronized a way as possible.

\subsection*{Acknowledgments}

The authors acknowledge partial support from CNPq, CAPES, and a FAPERJ BBP
grant.

\bibliography{gcs}
\bibliographystyle{plain}

\end{document}